\documentclass[twocolumn,preprintnumbers, amssymb,amsmath,aps,floatfix,nofootinbib,superscriptaddress,showpacs]{revtex4-1}
\usepackage{epsfig}
\usepackage{bm}
\usepackage{amssymb}
\usepackage{amsmath}
\usepackage{color}
\usepackage{subfigure}
\usepackage[colorlinks,
            linkcolor=blue,
            anchorcolor=black,
            citecolor=blue
            ]{hyperref}

\usepackage{soul}

\begin{document}

\title{Harmonics of Parton Saturation in Lepton-Jet Correlations at the EIC} 
\author{Xuan-Bo Tong}\email{tongxuanbo@cuhk.edu.cn} 
\affiliation{School of Science and Engineering, The Chinese University of Hong Kong, Shenzhen, Shenzhen, Guangdong, 518172, P.R. China}
\affiliation{University of Science and Technology of China, Hefei, Anhui, 230026, P.R.China}

\author{Bo-Wen Xiao}\email{xiaobowen@cuhk.edu.cn} 
\affiliation{School of Science and Engineering, The Chinese University of Hong Kong, Shenzhen, Shenzhen, Guangdong, 518172, P.R. China}

\author{Yuan-Yuan Zhang}
\email{zhangyuanyuan@cuhk.edu.cn} 
\affiliation{School of Science and Engineering, The Chinese University of Hong Kong, Shenzhen, Shenzhen, Guangdong, 518172, P.R. China}
\affiliation{University of Science and Technology of China, Hefei, Anhui, 230026, P.R.China}
 
\begin{abstract}
Parton saturation is one of the most intriguing phenomena in the high energy nuclear physics research frontier, especially in the upcoming era of the Electron-Ion Collider (EIC). The lepton-jet correlation in deep inelastic scattering provides us with a new gateway to the parton saturation at the EIC. In particular, we demonstrate that azimuthal angle anisotropies of the lepton-jet correlation are sensitive to the strength of the saturation momentum in the EIC kinematic region. In contrast to the predictions based on the collinear framework calculation, significant nuclear modification of the anisotropies is observed when we compare the saturation physics results in $e+p$ and $e+\text{Au}$ scatterings. By measuring these harmonic coefficients at the EIC, one can conduct quantitative analysis in different collisional systems and unveil compelling evidence for saturation effects.

\end{abstract}
\maketitle
%\fi
{\it Introduction.} The parton saturation phenomenon~\cite{Gribov:1983ivg,Mueller:1985wy,Mueller:1989st,McLerran:1993ni,McLerran:1993ka,McLerran:1994vd}, predicted by the color glass condensate~(CGC) theory~\cite{Gelis:2010nm,Iancu:2003xm}, is one of the cutting-edge topics that many physicists strive to explore, and it is also one of the profound questions that the electron-ion collider (EIC)~\cite{Boer:2011fh,Accardi:2012qut,AbdulKhalek:2021gbh,AbdulKhalek:2022hcn} intends to address. Various observables have been proposed to search for the signals of the parton saturation at the future EIC~(see e.g., a recent review ~\cite{Morreale:2021pnn}). Especially, two-particle correlations~\cite{Dominguez:2010xd,Dominguez:2011wm,Mueller:2013wwa,Metz:2011wb,Dominguez:2011br,Dumitru:2015gaa,Dumitru:2016jku,Boer:2016fqd,Dumitru:2018kuw,Mantysaari:2019hkq,Zhao:2021kae,Boussarie:2021ybe,Caucal:2021ent,Zhang:2021tcc,Taels:2022tza,Caucal:2022ulg,Boussarie:2014lxa,Boussarie:2016ogo,Salazar:2019ncp,Boussarie:2019ero,Boer:2021upt,Iancu:2021rup,Iancu:2022lcw,Hatta:2016dxp,Altinoluk:2015dpi,Mantysaari:2019csc,Hagiwara:2021xkf,Zheng:2014vka,Bergabo:2021woe,Bergabo:2022tcu,Iancu:2022gpw,Fucilla:2022wcg,Kolbe:2020tlq}, including the azimuthal correlations of di-jet~\cite{Dominguez:2010xd,Dominguez:2011wm,Mueller:2013wwa,Metz:2011wb,Dominguez:2011br,Dumitru:2015gaa,Dumitru:2016jku,Boer:2016fqd,Dumitru:2018kuw,Mantysaari:2019hkq,Zhao:2021kae,Boussarie:2021ybe,Caucal:2021ent,Zhang:2021tcc,Taels:2022tza,Caucal:2022ulg,Boussarie:2014lxa,Boussarie:2016ogo,Salazar:2019ncp,Boussarie:2019ero,Boer:2021upt,Iancu:2021rup,Iancu:2022lcw,Hatta:2016dxp,Altinoluk:2015dpi,Mantysaari:2019csc} or di-hadron~\cite{Zheng:2014vka,Bergabo:2021woe,Bergabo:2022tcu,Iancu:2022gpw,Fucilla:2022wcg} in the inclusive or diffractive processes, are one of the promising observables under intensive study in the last decade, see also the tremendous applications in the LHC and  RHIC~\cite{Jalilian-Marian:2004vhw,Kharzeev:2004bw,Marquet:2007vb,Tuchin:2009nf,Dumitru:2010ak,Dominguez:2010xd,Dominguez:2011wm,Mueller:2013wwa,Kutak:2012rf,vanHameren:2014ala,Kotko:2015ura,vanHameren:2016ftb,vanHameren:2019ysa,vanHameren:2020rqt,Marquet:2016cgx,Klein:2019qfb,Iancu:2020mos,Bolognino:2021mrc,Al-Mashad:2022zbq,Hagiwara:2017fye,Kotko:2017oxg,Bhattacharya:2018lgm,Boussarie:2018zwg,Albacete:2010pg,Stasto:2011ru,Lappi:2012nh,Iancu:2013dta,Albacete:2018ruq,Stasto:2018rci,Jalilian-Marian:2005qbq,Jalilian-Marian:2012wwi,Stasto:2012ru,Rezaeian:2012wa,Basso:2015pba,Rezaeian:2016szi,Basso:2016ulb,Boer:2017xpy,Benic:2017znu,Goncalves:2020tvh,Benic:2022ixp,Boer:2022njw,Gelis:2002fw,Kovner:2014qea,Kovner:2015rna,Marquet:2019ltn,Akcakaya:2012si,Marquet:2017xwy}.
The azimuthal angle anisotropies of two-particle correlations can be obtained through harmonic analysis, and they offer unique perspectives and channels to identify and probe the parton saturation at the  EIC~\cite{Metz:2011wb,Dominguez:2011br,Dumitru:2015gaa,Dumitru:2016jku,Dumitru:2018kuw,Boer:2016fqd,Mantysaari:2019hkq,Zhao:2021kae,Boussarie:2021ybe,Hatta:2016dxp,Altinoluk:2015dpi,Mantysaari:2019csc,Hagiwara:2021xkf}.

Recently, the lepton-jet correlation~(LJC), a novel type of two-particle correlation, was introduced in Refs.~\cite{Liu:2020dct,Liu:2018trl} to study the quark transverse-momentum-dependent (TMD) distributions in deep inelastic scatterings. It is interesting to note that this observable is directly defined in the lab frame (or the center-of-mass frame of the incoming lepton and hadron) and thus can be conveniently measured by the detector. This process allows one to use the scattered lepton as the tagging reference, which can be identified and measured rather precisely, and the recoiled jet as the probe~\cite{Arratia:2019vju,H1:2021wkz} to study the soft dynamics of the process in the back-to-back correlation region, where the system's imbalance transverse momentum $|\vec q_\perp|$ is much softer than its relative transverse momentum $|\vec P_\perp|$. Since the Bremsstrahlung emission of gluon (and photon) favors the collinear region, it was shown that large radiation-induced anisotropy in LJC~\cite{Hatta:2021jcd} can be generated within the TMD and collinear factorization framework  \cite{Hatta:2021jcd,Hatta:2020bgy,Catani:2014qha,Catani:2017tuc,Ju:2022wia}.

The objective of this paper is to use the LJC anisotropy as a novel probe to study parton saturation in detail. As an early exploration, we present quantitative evaluations of several LJC anisotropies with various inputs in the CGC and collinear formalisms. In particular, we demonstrate that the $ \langle \cos n \phi \rangle $ coefficients, namely, the harmonics of LJC, can shed light on distinct behaviors of the saturation phenomenon in small-$q_\perp$ region. Here $\phi$ is the azimuthal angle difference between $\vec q_{\perp}$ and $\vec P_{\perp}$, and $n$ is the harmonic number. 

First, due to the collinear enhancement of final-state jet gluon emission, the near-cone gluon emission has the largest probability in general, and thus produces significant initial anisotropies. Second, through numerical and analytic analysis, we find that the saturation effects can significantly suppress the anisotropy in the small $q_\perp$ region especially in $e+\textrm{Au}$ collisions. In the CGC framework, the typical transverse momentum of incoming quarks is related to and determined by the saturation momentum $Q_s$, and it is due to random multiple scatterings and small-$x$ gluon emissions, which have no angular preference. Thus, one expects that saturation effects tend to reduce the final measured LJC anisotropies. For the cases with appreciable saturation momentum $Q_s$, the anisotropy can be largely washed out in the region $|\vec q_\perp | \lesssim Q_s$. In addition, due to the nuclear enhancement of the saturation momentum $Q_s^2\propto   A^{1/3}$ with $A$ the nucleon number, a strong nuclear modification of the final anisotropies is then predicted by changing the target hadron from the proton to heavy nuclei, such as $\textrm{Au}$. Thus, this new observable at the upcoming EIC may yield additional compelling evidence for the parton saturation.

{\it Lepton-jet correlation at small $x$.} 
We investigate the production of the lepton and jet in deep inelastic scatterings as follows
\begin{align}
    \ell(k)+\textrm{A} (p)\rightarrow \ell^{\prime}\left(k_{\ell}\right)+\operatorname{Jet}\left(k\textbf{}_{J}\right)+X, 
\end{align}
and measure their azimuthal angular correlations in the back-to-back region ($|\vec q_{\perp}|=| \vec{k}_{\ell\perp} + \vec{k}_{J\perp}| \ll  |\vec{P}_{\perp} |= |(\vec{k}_{\ell\perp} - \vec{k}_{J\perp}) /2|$).  Based on the TMD factorization and the small-$x$ formalism~\cite{Marquet:2009ca,Dominguez:2010xd,Xiao:2010sa,Xiao:2010sp,Dominguez:2011wm}, we derive the azimuthal angle dependent cross section for LJC in the correlation limit. To include the parton showers effects, the Sudakov logarithms are resumed to all orders using the technique developed in Refs.~\cite{Mueller:2012uf,Mueller:2013wwa} and \cite{Hatta:2021jcd,Hatta:2020bgy,Catani:2014qha,Catani:2017tuc}. In terms of the harmonic expansion, the cross section can be cast into:
\begin{align} 
  & \frac{d^{5} \sigma(\ell P \rightarrow \ell^{\prime} J)}{d y_{\ell} d^{2} P_{ \perp} d^{2} q_{\perp}}=
\sigma_0\int  \frac{b_{\perp}db_{\perp}}{2\pi} W(x,b_\perp)  \Big[J_0( q_\perp b_\perp)
  \notag \\ 
  &  
 \qquad + 
\sum_{n=1}^\infty 2\cos( n\phi) \alpha_s(\mu_{b})\frac{ C_Fc_n(R)}{n\pi} J_n(q_\perp b_\perp) \Big] ~.
 \label{eq:diff}
 \end{align}
 Here we can follow the procedures prescribed in Refs.~\cite{Mueller:2012uf,Mueller:2013wwa}   and assume that the relevant small-$x$ logarithms can be resumed into the small-$x$ quark distributions~\cite{Xiao:2017yya,Chirilli:2011km,Chirilli:2012jd}. The rigorous demonstration of the factorization is expected to be similar to that for di-jet correlations, e.g.,~\cite{Caucal:2021ent,Taels:2022tza,Caucal:2022ulg}. 
 \par  
 At leading order (LO), the correlation is produced by a quark (or antiquark) jet and the away side lepton. In Eq.(\ref{eq:diff}), $x$ is the longitudinal momentum fraction of the incoming quark w.r.t. the target proton or nucleus. From kinematics, one finds $x =k_{\ell \perp}/\sqrt{s_{e N}}\left(e^{-y_{\ell}}+e^{-y_{J}}\right)$ and $1 =k_{\ell \perp}/\sqrt{s_{e N}}\left(e^{y_{\ell}}+e^{y_{J}}\right)$, where $y_\ell,\,y_J$ are the rapidities of the produced lepton and jet, respectively and $s_{eN}=(k+p)^2$ is the center-of-mass energy. $\sigma_0=({\alpha_e^2}/{\hat{s} Q^2}) [{2(\hat{s}^2+\hat{u}^2)}/{Q^4}]$ is the LO hard cross section with $\hat{s}$, $\hat{u}$ as Mandelstam variables of the partonic subprocess and $Q^2=-(k-k_\ell)^2$ as the lepton momentum transfer. $W$ function is defined as
\begin{align}
W(x,b_\perp)=  \sum_q e_q^2 xf_q(x,b_{\perp}) e^{-\text{Sud}(b_\perp)}~,
\label{eq:W}
\end{align}
where $f_q(x,b_{\perp})$ denotes the unintegrated quark distribution in the coordinate space.  In small-$x$ limit, it is related to the dipole scattering matrix as follows~\cite{McLerran:1998nk,Venugopalan:1999wu,Mueller:1999wm,Marquet:2009ca,Xiao:2017yya}
\begin{align}
x f_q(& x, b_{\perp})
= \frac{N_{c}S_\perp}{8\pi^{4}}   \int  d \epsilon_f^2 d^2 r_{\perp} 
\frac{(\vec b_\perp +\vec r_\perp) \cdot \vec r_\perp }{ | \vec b_\perp +\vec r_\perp||\vec r_\perp| } 
\notag \\ 
&\times \epsilon_f^2  K_{1}(\epsilon_f | \vec b_\perp +\vec r_\perp|) K_{1}\left(\epsilon_f |\vec r_{\perp}|\right) 
\notag \\ 
&\times \Big [1+{\cal S}_x(b_\perp )-{\cal S}_x(b_\perp+r_\perp)-{\cal S}_x(r_\perp)\Big  ]~,
\label{eq:match}
\end{align} 
where $S_\perp$ is the averaged transverse area of the target hadron and ${\cal S}_x(r_{\perp})$ represents the dipole scattering matrix with $r_{\perp}$ the dipole transverse size. 

The Sudakov form factor $\text{Sud}(b_\perp)$ in Eq.(\ref{eq:W})  
can be separated into the perturbative and non-perturbative parts, $\text{Sud}=\text{Sud}_{\text{P}}+\text{Sud}_{\text{NP}}$. The perturbative Sudakov factor reads
 \begin{equation}
 \text{Sud}_{\text{P}}=\int_{\mu_{b}}^{Q} \frac{d \mu}{\mu} \frac{\alpha_{s}(\mu) C_{F}}{\pi}
\Big[\ln \frac{Q^{2}}{\mu^{2}}
+\ln \frac{Q^{2}}{P_{\perp}^{2}}
 +c_{0}(R)\Big]~,
 \label{eq:sudper}
 \end{equation}
where $\mu_b=b_0/b_\perp^*$ with $b_0\equiv2e^{-\gamma_E}$ and $\gamma_E$ the Euler constant; $b^*_\perp=b_\perp/\sqrt{1+b_\perp^2/b_\text{max}^2}$ with $b_{\text{max}}=1.5~$GeV$^{-1}$~\cite{Sun:2014dqm}. Since the small-$x$ quark distributions have already contained the non-perturbative information on the target, $\text{Sud}_{\text{NP}}$ is usually not included in the CGC formalism for simplicity. For comparison purposes, we also compute the LJC in the TMD framework~\cite{Hatta:2021jcd} with the collinear quark distribution and the standard non-perturbative $\text{Sud}_{\text{NP}}$~\cite{Sun:2014dqm,Prokudin:2015ysa}.

The Fourier coefficient $c_n(R)$ is extracted by analyzing the soft gluon emission from the initial quark and final jet, following the procedure in Ref.~\cite{Hatta:2021jcd}. After subtracting soft gluon emissions inside jets, one obtains the residual dependence on the jet cone size $R$ in $c_n(R)$. The exact expression of $c_n(R)$ can be found in Ref.~\cite{Hatta:2021jcd}. In our calculation, we choose  $R=0.4$ , where $c_0=1.79,~c_1=2.60,~c_2=0.95,~c_3=0.25$.
\par

With the harmonic expansion of the differential cross section in Eq.(\ref{eq:diff}), the azimuthal anisotropy of the LJC reads 
\begin{equation}
\langle \cos n\phi \rangle 
  =\frac{\sigma_0  \int b_{\perp} d b_{\perp} J_{n}\left(q_{\perp}b_{\perp}\right) 
W(x,b_\perp) \alpha_s(\mu_{b})\frac{C_{F}  c_{n}(R)}{n \pi} 
 }
{ \sigma_0 \int b_{\perp} d b_{\perp} J_{0}\left(q_{\perp}b_{\perp}\right)  W(x,b_\perp)  }~.
\label{eq:azimuthal_asym}
\end{equation}
Using various models for the dipole S-matrix, we can predict the LJC azimuthal anisotropy $\langle \cos n\phi \rangle$ at small $x$, and show the dependence on the momentum imbalance $q_\perp$ and the sensitivity to the saturation momentum $Q_s$.

At last, let us discuss the QED modification to the anisotropy $\langle \cos n\phi \rangle$. As noted by Ref.~\cite{Hatta:2021jcd}, this modification can occur when we turn on the final-state photon emissions from the electron side. In this case, we find that the formula of $\langle \cos n\phi \rangle$ in Eq.(\ref{eq:azimuthal_asym}) can be modified by the following two corrections. First, the QED Sudakov form factor should be added, which reads 
\begin{align}
    \text{Sud}^\gamma =\int^Q_{b_0/b_\perp}\frac{d \mu}{\mu} \frac{\alpha_e}{\pi}\Big[\ln \frac{Q^2}{\mu^2}+\ln \frac{Q^2}{P_{\perp}^2}-\frac{3}{2}+\ln \frac{P_{\perp}^2}{m_e^2}\Big]
    \label{eq:sudperQED}
\end{align} with $m_e$ the electron mass and $\alpha_{e}$ the QED coupling.
Second, the Fourier coefficient $\alpha_s C_F c_n $ is replaced by $\alpha_s C_F c_n+\alpha_{e} c_{n}^\gamma$, where 
\begin{equation}
        c_{n}^\gamma=(-1)^n\Big [\ln \frac{P_\perp^2}{m_e^2}+\frac{2}{\pi}\int^\pi_0 d \phi (\pi- \phi)\frac{\cos \phi}{\sin \phi }(\cos n \phi-1)\Big]~.
    \label{eq:cgamma}
\end{equation}
We note that $c_{n}^\gamma$ is in line with $c_n(R)$ in the small-$R$ limit when $R$ is substituted with $m_e^2/P_\perp^2$, except for the overall factor $(-1)^n$. Moreover, despite the fact that the QED coupling $\alpha_{e}\approx 1/137$ is much smaller than that in QCD, the soft photon contribution to the anisotropy is enhanced by the large logarithm $\ln (P_\perp^2/m_e^2)$ in the correlation region.

\begin{figure*}[htpb]
\includegraphics[scale=0.84]{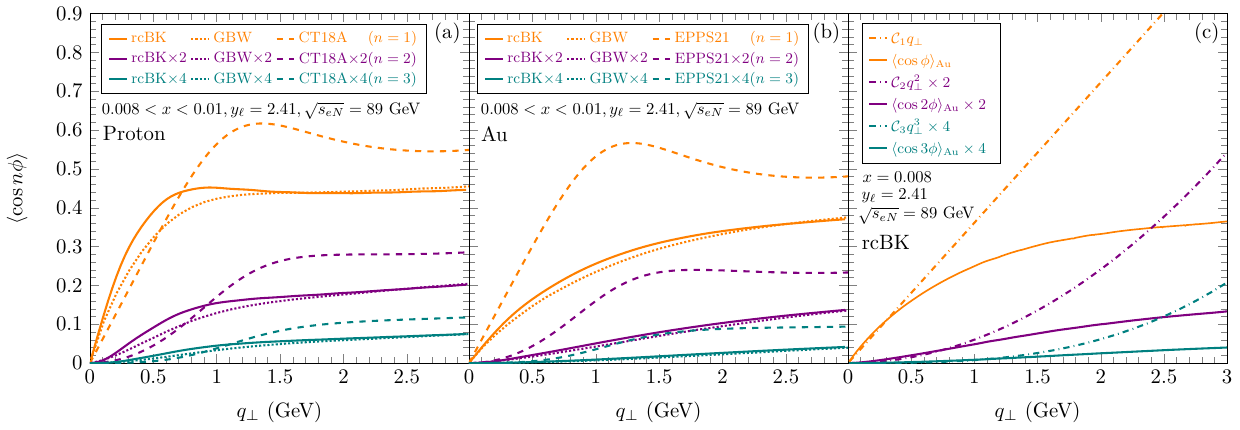}
 \caption{ (a) Harmonics in $e+p$ collisions with the inputs from the rcBK solution, GBW model, and CT18A PDFs; (b) Harmonics predicted for $e+\text{Au}$ collisions with the inputs: the rcBK solution, GBW model, and EPPS21 PDFs;  (c) Comparison of the exact results and the small-$q_\perp$ asymptotic behaviors. Solid lines: the exact results. Dashdotted lines: the small-$q_\perp$ asymptotic expansions in Eq.(\ref{eq:smallqT}).}  
      \label{fig:Asymmetry}
\end{figure*}

 {\it Numeric analysis } To numerically study the harmonic coefficients $\langle \cos n\phi \rangle $ in LJC, we employ the GBW model~\cite{Golec-Biernat:1998zce} and the solution of the running-coupling Balitsky-Kovchegov~(rcBK) equation~\cite{Balitsky:1995ub, Kovchegov:2006wf, Kovchegov:1999yj, Kovchegov:2006vj, Albacete:2010sy,Golec-Biernat:2001dqn,Albacete:2007yr,Balitsky:2006wa,Gardi:2006rp,Albacete:2007yr,Balitsky:2007feb,Berger:2010sh} for the dipole S-matrix ${\cal S}_x(r_{\perp})$. For the gold nucleus target,  the corresponding saturation momentum is chosen as $Q_{s,A}^2\approx 5Q_{s,p}^2$, where $Q_{s,p}$ is the proton saturation momentum. In the GBW model, ${\cal S}_x(r_{\perp})=e^{-r_{\perp}^2Q_s^2(x)/4}$,
 $Q_{s,p}^2(x)=(x_0/x)^{0.28}~\text{GeV}^2$ with $x_0 = 3\times 10^{-4}$. The modefied MV model~\cite{Fujii:2013gxa,Albacete:2010sy}, which gives ${\cal S}_{x_0}(r_{\perp})=\exp \left[-(r_{\perp}^{2} Q_{s 0}^{2}) ^{1.118}\ln (1/(0.24 r_{\perp})+e)/4\right]$ with $Q_{s0,p}^2=0.16$~GeV$^2$ at $x_0=0.01$, sets the initial conditions for the rcBK evolution. As a comparison, we also compute $\langle \cos n\phi \rangle $ in the TMD framework following Ref.~\cite{Hatta:2021jcd} with collinear parton distribution functions (PDFs) as inputs. Here we adopt the NLO PDF sets of CT18A~\cite{Hou:2019efy} and EPPS21~\cite{Eskola:2021nhw} parametrizations for proton and the gold nucleus, respectively.

With the above inputs, we predict the $q_\perp$-distribution of the $\langle \cos n\phi \rangle$ anisotropies for the following kinematics bin: $\sqrt{s_{eN}}=89$ GeV, $y_\ell=2.41$,\ $0.008\leq x\leq 0.01,\ 4$ GeV$\leq P_\perp\leq4.4$ GeV, 5.6~GeV$\leq Q\leq$ 5.9~GeV. This choice of kinematics coincides with the event simulation study of the LJC in Ref.~\cite{Arratia:2019vju} for the EIC. Due to limited collisional energies, it is difficult to access lower $x$ regions ($x\leq 1\times 10^{-3}$). Nevertheless, we later show that the saturation effects can be amplified with the gold nuclear target. Here the lower cut for $x$ is chosen as the minimum value allowed by the kinematic constraints at given $s_{\text{eN}},P_\perp$, namely $x_{\text{min}}\approx 4 P_\perp^2/s_{\textrm{eN}}$. 
 
  Fig.~\ref{fig:Asymmetry} shows $\langle \cos n\phi \rangle$ for $n=1,2,3$ as a function of $q_\perp $ in different models with the proton and gold nucleus targets. The results based on the rcBK solution and the GBW model are presented in the solid and dotted lines, respectively. Here the predictions are made at the jet cone size $R=0.4$.

  \begin{figure}
\centering
   \includegraphics[scale=0.9 ]{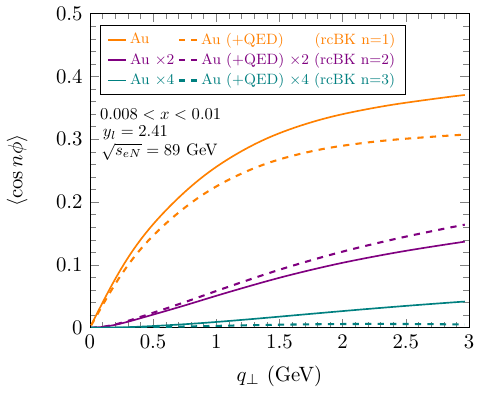}
   \caption{QED modifications to the LJC harmonics with the input from the rcBK solution for $e+\text{Au}$ collisions. Solid and dashed lines represent the harmonics without and with QED modifications, respectively.}
   \label{fig:AsymmetryQED}
\end{figure}
  
Let us comment on $\langle \cos n\phi \rangle$ shown in Figs.~\ref{fig:Asymmetry}~(a,b). The figures show that the anisotropies start from zero with sharp rises in the small-$q_\perp$ region, and then their increase becomes gradual in the larger-$q_\perp$ region. More specifically, we find that the anisotropies follows a power law at small-$q_\perp$, namely, $\langle \cos n\phi \rangle\sim q_\perp^n$, as in Fig.~\ref{fig:Asymmetry}~(c). The small-$q_\perp$ behaviour of the harmonics is an interesting tool for us to study the saturation effects in LJC. Moreover, the anisotropy decreases quickly with the harmonic number $n$. Since the dominant asymmetry arises due to the gluon emission from the quark jet side, the first harmonic ($n=1$) is the largest coefficient. Due to this hierarchy, we only plot the first three $n$. Nevertheless, our results show that sizable anisotropies can be measured at the EIC at least for harmonics with  $n=1, 2$. The above features are consistent with the collinear framework results, see e.g., the dashed lines in Figs.~\ref{fig:Asymmetry}~(a,b) and a previous calculation for proton in Ref.~\cite{Hatta:2021jcd}.

Furthermore, in Fig.~\ref{fig:AsymmetryQED}, we show the effect of the QED correction to $\langle \cos n\phi \rangle$ in dashed lines. Using the rcBK solution as the input, one can find that the QED corrections are fairly apparent. Especially, we find that the QED correction reduces the odd harmonics while it increases the even harmonics. This is expected since the QED Bremsstrahlung radiation favors the collinear region along the lepton direction, which is the away side of the gluon radiation. Analytically, this feature is reflected by the sign difference between the Fourier coefficient $c_{n}^\gamma$ and $c_n(R)$ for the odd harmonics, as discussed in Eq. (\ref{eq:cgamma}).

\begin{figure}
\centering
 \includegraphics[scale=0.9 ]{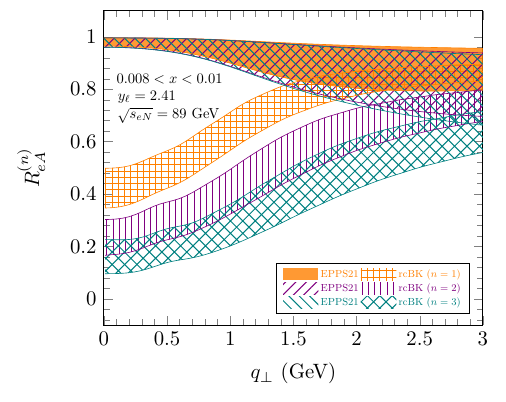}
   \caption{Nuclear modification factors of the harmonics. The upper bands represent ${R}_{eA}^{(n)}$ based on the inputs of the EPPS21 gold nuclear PDFs with uncertainties. The lower bands represent ${R}_{eA}^{(n)}$ calculated with the rcBK solution, where the gold saturation scale $Q_{s,A}^2$ varies from $3Q_{s,p}^2$ (upper bound in each band) to $5Q_{s,p}^2$ (lower bound).
   }
   \label{fig:ratioQED}
\end{figure}

In fact, by comparing the $q_\perp$-distributions of the harmonics between $e+p$ and $e + \text{Au}$ collisions in Figs.~\ref{fig:Asymmetry} (a) and (b), we observe a sizable decrease of the anisotropy $\langle \cos n\phi \rangle$ in the small-$x$ formalism  when switching from the proton target to the gold nucleus target. Since the saturation effects are isotropic in this process, the anisotropies are suppressed when the saturation momentum increases. To make this suppression more transparent, we define the nuclear modification factor as follows
\begin{align}
   {R}_{eA}^{(n)}= \frac{\langle \cos n\phi  \rangle_{eA}}{\langle \cos n\phi  \rangle_{ep}}~.
   \label{eq:ratioReA}
\end{align}
As shown in Fig.~\ref{fig:ratioQED}, significant nuclear suppression is predicted for all three harmonics in the low-$q_\perp$ region. On top of that, the saturation formalism indicates a numerical hierarchy of suppression as the harmonic number $n$ increases. This implies that the higher harmonics are more sensitive to the parton saturation effects. 

Also, we find that the QED modifications to the ratio ${ R}_{eA}^{(n)}$ in Eq.~(\ref{eq:ratioReA}) are negligible. 
This observation can be understood as follows. For one thing, although the QED correction to the Fourier coefficient, $\alpha_{e}c_{n}^\gamma$, can have apparent contributions to the harmonics as shown in Fig.~\ref{fig:AsymmetryQED}, it cancels between $\langle \cos n\phi  \rangle_{eA}$ and $\langle \cos n\phi  \rangle_{ep}$ in the ratio ${ R}_{eA}^{(n)}$; For another, the Sudakov form factor is dominated by the QCD contributions, which can be checked in Eqs. (\ref{eq:sudper},\ref{eq:sudperQED}).

    In addition, one can conduct a comparative study by computing the above nuclear modification factor ${R}_{eA}^{(n)}$ based on the inputs of the nuclear PDFs. Within the collinear framework, one can encode various nuclear effects in the nuclear PDFs by fitting them with existing experimental data. The ratios $R^{(n)}_{eA}$  calculated with the EPPS21 gold nuclear PDFs~\cite{Eskola:2021nhw} and the CT18A proton PDFs~\cite{Hou:2019efy} are presented in Fig.~\ref{fig:ratioQED} with the error band at $90\%$ confidence-level. Here we neglect the uncertainty from the baseline proton PDFs, since they are small. Interestingly, we find that the predictions of $R^{(n)}_{eA}$ made with the rcBK solution and EPPS21 PDFs are substantially different at small $q_\perp$, although the predictions at large $q_\perp$ all start to converge toward unity.

The nuclear modification factor plotted in Fig.~\ref{fig:ratioQED} is the main result of this work, since it illustrates the striking difference between the saturation and non-saturation frameworks.  Therefore, by measuring the LJC anisotropy in $e+p$ to $e+\text{Au}$ collisions, one can quantitatively study the parton saturation and pinpoint its signature features at the upcoming EIC.

{\it Analytic explanations of the numerical results.  } Given the anisotropies derived in correlation limit in Eq.~(\ref{eq:azimuthal_asym}), we further expand the expression in the small-$q_{\perp}$ limit. Since saturation effects are expected to be dominant in the small-$q_\perp$ region, this asymptotic expansion can provide an analytic understanding of the nuclear suppression of the harmonics due to parton saturation.
 
By approximating the Bessel functions $J_{n}\left(q_{\perp}b_{\perp}\right) \sim {(q_\perp b_\perp/2)^{n}}/{\Gamma(n+1)}$ in Eq.~(\ref{eq:azimuthal_asym}), we can obtain the simple power law behavior for the harmonics, which reads $\langle \cos n\phi \rangle\approx {\cal C}_n q_\perp^{n}$ with ${\cal C}_n$ defined as the ratio of two integrals over $b_\perp$. In the correlation limit $Q\gtrsim P_\perp\gg q_\perp $, one can justify the saddle-point approximation (see e.g., Refs.~\cite{Parisi:1979se,Collins:1981va,Collins:1984kg,Shi:2021hwx}) in the evaluation of these two integrals, and thus obtain the harmonics at small $q_\perp$
\begin{align}
\langle \cos n\phi \rangle
 &\approx  {\cal C}_n q_\perp^n
 \notag \\ 
  & \approx 
\left(\frac{q_{\perp} b_0 }{2\Lambda_{\text{QCD}}} \right)^n  
 \frac{\alpha_s(\mu^{\text{sp}}_n) C_F  c_{n} (R)}{\pi   n \Gamma(n+1)}\frac{%\ \sum_{q} e_q^2
  f_q(x,b^{\text{sp}}_{\perp n})  }
{%\sum_{q} e_q^2 \ 
 f_q(x, b^{\text{sp}}_{\perp 0})  
}~
\notag \\ &
\times  \left[\frac{ 2\beta_1+ C_F}{ (n+2)\beta_1+ C_F }  \right]
^{1+\frac{ C_F}{2\beta_{1}}\ln  \frac{e^{c_0(R)} Q^4}{ \Lambda_{\text{QCD}}^2 P_\perp^2} }
~,
%\notag \\ &
% \times 
\label{eq:smallqT}
\end{align}
where $\beta_1=(33-2 n_f)/12$, $\mu^{\text{sp}}_n= b_0/b^{\text{sp}}_{\perp n}$, $ b_0\equiv 2e^{-\gamma_E}$. $b^{\text{sp}}_{\perp n}$ denotes the position of the saddle point:
\begin{align}
b^{\text{sp}}_{\perp n} =\frac{ b_0}{\Lambda_{\text{QCD}} } \Bigg[\frac{  e^{c_0(R)}Q^4}{\Lambda_{\text{QCD}}^2 P_\perp^2 }\Bigg]^{-\frac{C_F}{2(2+n)\beta_1+2C_F}}~,
\label{eq:saddleP}
\end{align}
which is mainly determined by the Sudakov factor in Eq.(\ref{eq:sudper}). In particular, Eq.(\ref{eq:saddleP}) in the $n=0$ case is akin to the saddle point originally found in Ref.~\cite{Collins:1984kg}.

  Fig.~\ref{fig:Asymmetry}(c) presents the numerical comparison between the small-$q_\perp$ expansion of $\langle \cos n\phi \rangle$ in Eq.~(\ref{eq:smallqT}) and the exact expression in Eq.~(\ref{eq:azimuthal_asym}) with the rcBK solution as the input for the gold target. One can see that the two results coincide in the small-$q_\perp$ region.
 \par  
 Let us discuss the dependence of the saturation momentum $Q_s$ in the harmonics, which has been entirely contained in the ratio of unintegrated quark distribution $f_q(x,b_\perp^{\text{sp}})$ in Eq.~(\ref{eq:smallqT}).  One finds that $b^{\text{sp}}_{\perp n} \rightarrow 0$ in the limit $Q \gtrsim P_{\perp} \rightarrow \infty$ from Eq.~(\ref{eq:saddleP}). Thus we can apply small-$b_\perp$ approximation on $f_q(x,b_\perp^{\text{sp}})$, which yields $f_q(x,b_\perp^{\text{sp}})\propto Q_s^2\ln [1/(Q_s b_\perp^{\text{sp}})]$~\cite{Mueller:1999wm,Xiao:2017yya}. As a result, the anisotropic harmonics has the following asymptotic form,
\begin{align}
 \langle \cos n\phi \rangle  \propto   \frac{
f_q(x,b^{\text{sp}}_{\perp n})  }
{  f_q(x, b^{\text{sp}}_{\perp 0}) }\approx\frac{\ln(Q_s b^{\text{sp}}_{\perp n})  }{\ln(Q_s b^{\text{sp}}_{\perp 0})}~.
\label{eq:ratio}
\end{align}
Since Eq.~(\ref{eq:saddleP}) implies $b^{\text{sp}}_{\perp n}>b^{\text{sp}}_{\perp 0}$ for positive $n$, one can check the derivative of $\langle \cos n\phi \rangle$ with respective to $Q_s$ is negative. That means as the saturation momentum $Q_s$ increases, the anisotropic harmonics $\langle \cos n\phi \rangle$ decrease. Therefore, we can qualitatively explain the suppression of the anisotropy observed in the numerical study.
As illustrated in Fig. \ref{fig:ratioQED}, there is a numerical hierarchy of the nuclear modifications, which the nuclear suppression becomes stronger as the harmonic number $n$ increases. This can be qualitatively explained by Eq.~(\ref{eq:ratio}) and the fact that $b^{\text{sp}}_{\perp n}$ increases with $n$ according to Eq.~(\ref{eq:saddleP}).

{\it  Conclusion } 
In summary, by investigating the azimuthal angle anisotropy of the LJC in DIS in the correlation limit, we find that the LJC anisotropy is sensitive to saturation effects and thus provides us a novel channel to study the onset of the saturation phenomenon at the EIC. Through a comprehensive numerical study, we demonstrate that there is a considerable suppression in the anisotropy when the target changes from proton to the gold nucleus. This nuclear suppression is significant in the small-$q_\perp$ region, where the saturation effects are dominant. In contrast, we find much smaller nuclear suppression in the collinear factorization framework. Therefore, the nuclear modification factor $R_{eA}^{(n)}$ can be used as a benchmark to distinguish the saturation and non-saturation frameworks.
\par 

{\bf Acknowledgments:} 
We thank Feng Yuan for useful comments. This work is supported by the CUHK-Shenzhen university development fund under grant No. UDF01001859.

\newpage
\section*{SUPPLEMENTAL MATERIAL}
As the supplemental material, we outline the derivation of the harmonic expansion of the lepton-jet correlation~(LJC).
\par 
At leading order of $\alpha_s$, the LJC in the high energy $\ell A$ collisions comes from the exchange of a virtual photon between the initial lepton and a small-$x$ quark (or antiquark) from the target. In the correlation limit $|q_\perp|\ll |P_\perp|<Q$,  the LO differential cross-section of LJC can be factorized as~\cite{Liu:2018trl,Liu:2020dct}: 
\begin{equation}
\frac{d^5\sigma ^{(0)}}{d y_l d^2P_{\perp} d^2 q_{\perp}}= \sigma_0 \int d^2 v_{\perp} \delta^{(2)}(q_{\perp}-v_{\perp}) \sum_q e_q^2 x f_q(x,v_{\perp} )~% \notag \\ 
\end{equation}
where $\sigma_0 = ({\alpha_e^2}/{\hat{s} Q^2}) [{2(\hat{s}^2+\hat{u}^2)}/{Q^4}]$ with $\hat{s}$, $\hat{u}$ as Mandelstam variables of the partonic subprocess, and  
$f_q(x,v_{\perp} )$ is the small-$x$ unintegrated quark distribution~\cite{McLerran:1998nk,Venugopalan:1999wu,Mueller:1999wm,Marquet:2009ca,Xiao:2017yya}. 

Since the quark transverse momentum $v_\perp$ is isotropic, we need to consider additional soft-gluon emissions to obtain non-zero LJC azimuthal asymmetry in the partonic subprocess~\cite{Hatta:2020bgy,Hatta:2021jcd}. For one gluon radiation, the cross section can be written as 
\begin{equation}
\begin{aligned}
&\frac{d^5\sigma ^{(1)}}{d y_l d^2P_{\perp} d^2 q_{\perp}}= \sigma_0 \int  d^2v_{\perp}  \sum_q e_q^2xf_q(x,v_{\perp}) \\
&\quad\times \int d^2 k_{g\perp}S(k_{g\perp})\delta^{(2)}(q_\perp+k_{g\perp}-v_{\perp})~,
\end{aligned}
\end{equation}
where $S(k_{g\perp})$ is the standard eikonal formula that represents the total probability of one gluon emitted from an initial quark in the target and the final jet outside the jet cone:  
\begin{align}
S(k_{g\perp})=&g^2 C_F \int \frac{d y_g }{2(2\pi)^3}\frac{2k_J\cdot k_q }{k_J\cdot k_g \ k_q\cdot k_g}~.
\end{align}
Here $k_q,k_J$ denote the momenta of the incoming quark and final jet, respectively, and the constraint $\Delta_{k_g k_J}=(y_g-y_J)+(\phi_g-\phi_J)^2>R^2$ is implied. $y_{g},y_{J}$ and $\phi_{g},\phi_{J}$ are the rapidities and azimuthal angles of gluon and jet, respectively. %The details of the $y_g$-integration can be found in Ref.~\cite{Hatta:2021jcd}. 
\par 
To compute the azimuthal anisotropy, we perform the harmonic expansion with respect to the relative angle $(\phi_g-\phi_J)$ for the eikonal formula in Eq.~(\ref{eq:onegluon})~\cite{Hatta:2021jcd}. As a result, we write $S(k_{g\perp})=S_{\mathrm{iso}}(k_{g\perp})+S_{\mathrm{aniso}}(k_{g\perp})$ with the isotropic and anisotropic parts as follows
\begin{equation}
\begin{aligned}
S_{\mathrm{iso}}(k_{g\perp}) =& \frac{\alpha_s C_F}{2\pi^2 k_{g\perp}^2}\Big[\ln\frac{Q^2}{k_{g\perp}^2}+ \ln\frac{Q^2}{k_{J\perp}^2}+ c_0(R) \Big]~,\\
S_{\mathrm{aniso}}(k_{g\perp}) =& \frac{\alpha_s C_F}{2\pi^2 k_{g\perp}^2} 2 \sum_{n=1}^{\infty} c_n(R) \cos n(\phi_g-\phi_J) ~.
\label{eq:onegluon}
\end{aligned}
\end{equation}
The expression of the harmonic coefficient $c_n(R)$~\cite{Hatta:2021jcd} for arbitrary $R$ reads
\begin{align}
c_n(R)= & \frac{2}{\pi} \int_0^R d \phi \left\lbrace \frac{\cos \phi}{\sin \phi}\left[(\pi-\phi)-\tan ^{-1}\left(\frac{e^{y_+}-\cos \phi}{\sin \phi}\right) \right.\right. \notag \\ 
&\left. \left. +\tan ^{-1}\left(\frac{e^{y_-}-\cos \phi}{\sin \phi}\right)\right]   - y_+\right\rbrace\cos n\phi   \notag  \\ 
& +\frac{2}{\pi} \int_R^\pi d \phi \frac{\cos \phi}{\sin \phi}(\pi-\phi) \cos n \phi~,
\end{align} 
where $y_{\pm} = \pm \sqrt{R^2-\phi^2}$.
\par
Now let us consider the contributions from the soft gluon emissions to all order. The resummation of these contribution can be conveniently done by Fourier transform to the coordinate ($b_{\perp}$) space:
\begin{align}
\frac{d^5\sigma }{d y_l d^2P_{\perp} d^2 q_{\perp}}
\approx&\sigma_0 \int \frac{d^2b_\perp}{(2\pi)^2} e^{i {q}_{\perp}\cdot b_{\perp}} \sum_q e_q^2xf_q(x,b_{\perp})  
\notag \\ 
&\times e^{S_{\mathrm{iso}}(b_{\perp})} \Big[1+ S_{\mathrm{aniso}}(b_{\perp})\Big]~.
\label{eq:app1}
\end{align}
From Eq.~(\ref{eq:onegluon}), the isotropic part in $b$-space contains single and double logarithms, which needs to be resummed for reliable prediction. As a result,
the isotropic part yields the Sudakov form factor,
\begin{align}
S_{\mathrm{iso}}(b_{\perp})
=& - \int_{\mu_b}^{Q}\frac{d\mu}{\mu} \frac{\alpha_sC_F}{\pi} \left[ \ln \frac{Q^{2}}{\mu^{2}}+\ln\frac{Q^2}{P_{\perp}^2} + c_0(R)\right] 
\notag \\ 
=&-\text{Sud}_{\text{P}}~.
\label{eq:app2}
\end{align}
Here we have added the contributions from virtual diagrams to cancel the infrared divergences. Note that in the correlation limit, $P_\perp=(k_{\ell\perp}-k_{J_\perp})/2 \approx -k_{J_\perp}$. 

Compared to the isotropic contributions, the Fourier transformation of the anisotropic part does not have IR divergence. The convergence of the anisotropic part at $k_{g\perp}\rightarrow 0$ can be easily verified through Fourier transform. As a result, the Fourier coefficients of higher harmonics ($n>0$) vanish when the momentum imbalance $q_\perp$ goes to $0$.  Also, since this part is numerically small for a reasonably large cone-size $R$, thus we only keep the leading $\alpha_s$ order contribution. One can apply the expansion $e^{i z \cos (\phi)}=J_0(z)+2 \sum_{n=1}^{\infty} i^n J_n(z) \cos (n \phi) $
and the formula 
$
\int_0^{\infty} ({d z }/{z})J_n\left(z \left|b_{\perp}\right|\right)={1}/{n}
$
to perform the Fourier transform. It turns out that $S_{\mathrm{aniso}}$ have the following simple form:
\begin{equation}
\begin{aligned}
S_{\mathrm{aniso}}(b_{\perp})=\frac{\alpha_s C_F}{\pi} \sum_n(-i)^n c_n  \frac{2\cos n\phi_b }{n}
\end{aligned}
\label{eq:app3}
\end{equation}
where $\phi_b$ is the azimuthal angle of $b_\perp$ relative to $P_{\perp}$. 
\par 
With Eqs.~(\ref{eq:app1},\ref{eq:app2},\ref{eq:app3}), it is straightforward to obtain the harmonic expansion of the LJC in the correlation limit
\begin{equation}
\begin{aligned}
  & \frac{d^{5} \sigma(\ell P \rightarrow \ell^{\prime} J)}{d y_{\ell} d^{2} P_{ \perp} d^{2} q_{\perp}}=
\sigma_0\int  \frac{b_{\perp}db_{\perp}}{2\pi}\sum_q e_q^2xf_q(x,b_{\perp})  e^{-\text{Sud}_\text{P}} 
  \notag \\ 
  &  
 \Big[J_0( q_\perp b_\perp) + 
\sum_{n=1}^\infty 2\cos( n\phi) \alpha_s \frac{ C_Fc_n(R)}{n\pi} J_n(q_\perp b_\perp) \Big] ~,
\end{aligned}
\end{equation}
where $\phi$ is the azimuthal angle between $q_\perp$ and $P_{\perp}$.

In deriving the above results, we have assumed the small-$x$ factorization for the back-to-back lepton-jet productions. The more rigorous demonstration of the small-$x$ factorization of the lepton-jet correlation should involve the subtraction of the rapidity divergences and the cancellation of the infrared divergences. The rapidity divergence arises from the kinematic region where the radiated gluon is collinear to the target, and it can be absorbed into the small-$x$ distribution. After canceling the infrared divergence between real and virtual diagrams, one can obtain the Sudakov type logarithms for this process. 

Based on the previous one-loop studies~\cite{Chirilli:2011km,Chirilli:2012jd,Caucal:2021ent,Taels:2022tza,Caucal:2022ulg} in small-$x$ framework, the rapidity divergence can be subtracted from real and virtual diagrams and renormalized into the small-$x$ distribution. For the real diagram, one has to use the kinematic constraint and put in the proper energy cutoff, as shown in Ref.~\cite{Mueller:2012uf,Mueller:2013wwa}. In this procedure, the corresponding small-$x$ evolution equation can be re-derived, and the renormalization of the rapidity divergence is interpreted as the resummation of small-$x$ logarithms. We expect that the same procedure can be applied to the full one-loop calculation of the lepton-jet correlation as well.

After the rapidity subtraction, the infrared divergences left in the real and virtual diagrams should cancel, as shown in Ref.~\cite{Mueller:2012uf,Mueller:2013wwa}. In the end, one obtains the Sudakov logarithms due to the incomplete cancellation of the real and virtual contributions in the finite part. The Sudakov logarithms are the dominant leading power contribution when $q_{\perp} \ll P_{\perp}$ at the one-loop level. By performing the Sudakov resummation in $b_\perp$-space, one can achieve better predictive power and more reliable results. For the lepton-jet correlation, the technique for the derivation of the Sudakov factor is expected to be the same as Refs.\cite{Mueller:2012uf,Mueller:2013wwa,Catani:2017tuc,Hatta:2020bgy,Hatta:2021jcd}. As shown above, the Sudakov factor yields the isotropic part. In comparison, the anisotropic part, which gives rise to the harmonics, does not contain infrared divergences.

\end{document}